\documentclass[12pt,a4paper]{article}
\usepackage[T1]{fontenc}
\usepackage[utf8]{inputenc}
\usepackage[british]{babel}
\usepackage{amssymb}
\usepackage{amsmath}
\usepackage{mathrsfs}
\usepackage{bbm}
\usepackage[bookmarks=true,hyperfigures=true]{hyperref}
\usepackage[nosort]{cite}

\newcommand{\alg}[1]{\mathfrak{#1}}

\newcommand{\half}{\frac{1}{2}}
\newcommand{\thalf}{\tfrac{1}{2}}

\newcommand{\pslNN}{\ensuremath{\alg{psl}(N|N)}}
\newcommand{\psltt}{\ensuremath{\alg{psl}(2|2)}}
\newcommand{\sltt}{\ensuremath{\alg{sl}(2)\oplus\alg{sl}(2)}}
\newcommand{\g}{\alg{g}}
\newcommand{\gz}{\alg{g}^{(0)}}
\newcommand{\CPot}{\ensuremath{\mathbb{CP}^{1|2}}}
\newcommand{\ospft}{\ensuremath{\mathfrak{osp}(4|2)}}

\newcommand{\Cas}{\text{Cas}}

\newcommand{\comment}[1]{}

\author{Alessandra~Cagnazzo, Volker~Schomerus, Václav~Tlapák}
\date{\today}
\title{\bf High-Gradient Operators  \\[4mm] in the \psltt\ Gross-Neveu Model}

\begin{document}
\addtolength{\baselineskip}{5pt}

\maketitle
\begin{abstract}
It has been observed more than 25 years ago that sigma model
perturbation theory suffers from strongly RG-relevant high-gradient
operators. The phenomenon was first seen in 1-loop calculations
for the O(N) vector model and it is known to persist at least to two
loops. More recently, Ryu et al. suggested that a certain deformation
of the \pslNN\ WZNW-model at level $k=1$, or equivalently the \pslNN\
Gross-Neveu model, could be free of RG-relevant high-gradient operators
and they tested their suggestion to leading order in perturbation theory.
In this note we establish the absence of strongly RG-relevant high-gradient
operators in the \psltt\ Gross-Neveu model to all loops. In addition, we
determine the spectrum for a large subsector of the model at infinite coupling 
and observe that all scaling weights become half-integer. Evidence for a 
conjectured relation with the $\CPot$ sigma model is not found.
\end{abstract}

\vspace*{-15cm} {\flushright DESY-14-162}  \vspace*{15cm}

\section{Introduction}
\label{sec:introduction}
Non-linear sigma models (NLSM), with and without WZ term, play an important 
role in the description of condensed matter systems as well as string 
compactifications.
It has long been known \cite{Kravtsov:1989qs} that NLSM generically
suffer from the strong RG-relevance of certain high-gradient
operators.  At zero coupling the dimension of an operator in a NLSM
is determined by the number of derivatives, making high-gradient
operators highly irrelevant.  The usual assumption in perturbation
theory is that corrections to the scaling weight, the anomalous
dimensions, remain small as long as the
coupling does. However, it has been shown \cite{Kravtsov:1989qs} that this
assumption fails to hold for certain invariant high-gradient operators in the
$O(N)$-vector model and that in fact these operators can become relevant
even at infinitesimal values of the coupling. Similar results hold for
NLSM defined on a wide variety of compact and non-compact target 
(super-)spaces, see \cite{Ryu:2010iq} and references therein.

The generation of relevant operators represents a puzzling instability of the
UV fixed point. One might hope that higher orders in perturbation theory 
correct this issue. But unfortunately, the problem has been shown 
\cite{Castilla:1996qn} to become even worse at two loops. On the other hand, 
lattice simulations and other approaches have never shown signs of an 
instability, which adds to the long-standing puzzle. In the existing 
literature on the subject, the predominant attitude is to consider the 
RG-relevance of high-gradient operators as a bug, though Polyakov has  
argued \cite{Polyakov:2005ss} that it could be a desirable feature in order 
to establish a general pattern of dualities between NLSMs and WZNW models. 
In any case, the question remains whether the strong RG-relevance of 
high-gradient operators is corrected by higher loop or non-perturbative effects.  

The related WZNW models, when perturbed by a current-current
deformation, suffer from a similar issue \cite{Ryu:2010iq}, even if
the perturbation preserves conformal symmetry as it does for a number
of target supergroups. A few years ago, Ryu et al. suggested \cite{Ryu:2010iq}
that WZNW-models with affine \pslNN{} symmetry at level $k=1$ could be free
from strongly RG-relevant high-gradient operators. They also tested this
idea to leading order in perturbation theory, at least for certain classes
 of high-gradient operators.
In this note, we will exploit previously obtained results on the perturbation
theory of WZNW models and the representation theory of \psltt{} to prove that
the deformed \psltt\ WZNW model at $k=1$ is indeed free of strongly RG-relevant
\psltt\ invariant operators in any order of perturbation theory. In addition, we
evaluate the spectrum of all fields that transform in maximally atypical
(``$\frac12$ BPS'') representations of the target space symmetry \psltt\  up
to scaling weight $\Delta \leq 5$. Very remarkably, the spectrum at infinite
coupling turns out to assume half-integer values only, nurturing hopes it might
be described by a dual free field theory. In fact, it has been argued that
such a dual model is provided by the $\CPot$ NLSM \cite{CPdual}. Our results, 
however, do not provide evidence for a duality with this sigma model.

The plan of this short note is as follows. In the next section we
review some relevant results on all-loop anomalous dimensions in
perturbed WZNW models from \cite{Candu:2012xc}. In order to apply
these to the \psltt\ Gross-Neveu model, we need some background
from representation theory which is collected in section
\ref{sec:rep_theory}.
All-loop stability is established in section
\ref{sec:absence_of_high_gradient} before we compute
the low lying spectrum at infinite coupling in section
\ref{sec:spec_at_inf_g}.

\section{Results from WZNW perturbation theory}
\label{sec:pert_theory}

In this section we review recent results \cite{Candu:2012xc} on the
anomalous dimensions of WZNW models with Lie superalgebra symmetry.
If the Lie superalgebra $\g$ has vanishing dual Coxeter number
$g^{\vee}=0$, the current-current interaction
\begin{equation}
	\Omega(z,\bar{z})=J^{\mu}(z)\bar{J}_{\mu}(\bar{z})
	\label{eqn:interaction}
\end{equation}
is exactly marginal. Here, the index $\mu$ labels basis elements of
the Lie superalgebra $\g$ and we sum over all its allowed values.  The interaction
\eqref{eqn:interaction} clearly breaks the affine $\hat{\g}$-symmetry.
Moreover, it also breaks the chiral $\g$-symmetries, since the
operator $\Omega$ does not commute with the zero modes $J^{a}_{0}$
and $\bar{J}^{b}_{0}$ of the chiral currents. However, $\Omega$ does
commute with the sum of the zero modes $J^{a}_{0}+\bar{J}^{a}_{0}$,
thereby leaving the diagonal $\g$-symmetry intact.

The authors of \cite{Candu:2012xc} were able to obtain an all-order result
for the anomalous dimension of special operators in such $\Omega$
perturbed conformal field theories. It applies to all operators that
transform in a maximally atypical representation under the diagonal
action of the superalgebra $\g$. Maximally atypical (or $\frac12$BPS)
representations are indecomposables that contain a subrepresentation
with non-vanishing superdimension. The anomalous dimension $\delta_g$ 
for such
operators turns out to only depend on the representation labels and
the level $k$ of the affine superalgebra,
\begin{equation}
	\delta_g  = \frac{g}{2(1-k^{2}g^{2})}\bigl(\Cas^D -
	(1-kg) (\Cas^L + \Cas^R)\bigr),
	\label{eqn:anom_dim}
\end{equation}
where $\Cas$ is the quadratic Casimir operator and the superscripts
$D$, $L$ and $R$ indicate the diagonal, left and right action of the
algebra, respectively.

Let us now specialize to the case $\g=\psltt$. The superalgebra $\psltt$
has only one atypicality condition and the quadratic Casimir vanishes on
all atypical representations. Thus, eq.\ \eqref{eqn:anom_dim} simplifies
to
\begin{equation}
	\delta_g = -\frac{g}{2(1+kg)} (\Cas^L + \Cas^R).
	\label{eqn:anom_dim_psl}
\end{equation}
We are particularly interested in operators that are invariant under
the diagonal action of the symmetry algebra since such operators could
be used to generate a $\g$ preserving perturbation. The assumption of
$\g$ invariance does not simplify our formula \eqref{eqn:anom_dim_psl}
any further but it restricts it to operators for which the tensor
product of left and right action contains the trivial representation.
The finite-dimensional representation theory of $\psltt$ has been
worked out in detail in \cite{Gotz:2005ka}. The results imply that the
only way to obtain an invariant with $\Lambda_R = \Lambda$ is to
tensor with the same representation $\Lambda_L = \Lambda$.

\section{Review of \texorpdfstring{$\psltt$}{psl(2|2)}\ representation theory}
\label{sec:rep_theory}

In this section we give a brief review of the pertinent facts
regarding the Lie superalgebra $\psltt$ and its finite dimensional
representation theory. The algebra $\psltt$ has rank two and its
even subalgebra is $\g^{(0)}\simeq\sltt$. Consequently, all finite
dimensional representations are uniquely characterized by a pair of
$\alg{sl}(2)$ weights $j,l\in\half\mathbb{Z}$. Representations of
$\psltt$ can satisfy one shortening, or atypicality, condition which
is simply given by $j=l$. We will denote typical representations of
$\psltt$ by $[j,l]$ and atypical irreducibles by $[j]$. Irreducible
representations of the even subalgebra will be denoted by $(j,l)$.

Upon restriction to the even subalgebra $\g^{(0)}$ the irreducible
representations decompose as
\begin{align}
	[j]\big|_{\gz} &\simeq (j+\thalf,j-\thalf)\oplus
	2(j,j)\oplus(j-\thalf,j+\thalf)
	\qquad \text{(for $j>0$)}
	\label{eqn:atypical_rep_restriction}
	\\[2mm]
	[j,l]\big|_{\gz} &\simeq (j,l)\otimes\bigl[
	2(0,0)\oplus 2(\thalf,\thalf) \oplus (0,1) \oplus (1,0)
	\bigr]
	\label{eqn:typical_rep_restriction}
\end{align}
and $[0]$ is the trivial representation. Let us also remark that
$[\half]$ corresponds to the adjoint representation. Atypical irreducibles
of Lie superalgebras can form indecomposables. If one is not interested in
the precise form in which such indecomposables are built from their
constituents, all tensor products of finite dimensional $\g$
representations may be determined by restricting the factors to the
even subalgebra, tensoring the associated $\g^{(0)}$ representations
and combining the resulting products back into representations of
$\g$. The first and last step require no more than our decomposition
formulas \eqref{eqn:atypical_rep_restriction} and
\eqref{eqn:typical_rep_restriction}. The tensor products of
irreducibles, including the indecomposable structures, have been
worked out in \cite{Gotz:2005ka}.

We will also need to know the eigenvalue of the quadratic Casimir
invariant $\Cas$. It is given in terms of the highest weights by
\begin{equation}
	\begin{split}
		\Cas\bigl([j,l]\bigr) &= -j(j+1) + l(l+1)
		\\[2mm]
		\Cas\bigl([j]\bigr) &= 0.
	\end{split}
	\label{eqn:casimir}
\end{equation}
Note that its value in atypical representations is given by evaluating
the Casimir for typicals on weights which satisfy the shortening
condition $j=l$.

Let us conclude with a few scattered comments on the notation we are
about to use. As we mentioned above, atypical irreducibles can combine
to form complicated indecomposables. We will not concern ourselves with
this indecomposable structure of the spectrum and simply look at the
constituent irreducible representations. For this reason, we shall not
use the symbol $\oplus$ in our formulas but simply write $+$ instead.
Many of the sums of representations we are about to see are in fact
not direct. Since traces are blind to the indecomposable structures,
our formulas for representations encode true identities among their
characters $\chi_\Lambda$ in which $+$ and tensor products are
ordinary sums and products of characters.

\section{Absence of relevant high-gradient operators}
\label{sec:absence_of_high_gradient}

The spectrum of WZNW models on type I supergroups is quite well
understood, see \cite{Quella:2007hr}. Almost all of these models give
rise to logarithmic conformal field theories, see also
\cite{Schomerus:2005bf,Quella:2013rev}, and hence their Hamiltonian 
(generator of dilations) is not diagonalizable. In our analysis of the 
spectrum we shall only be concerned with the generalized eigenvalues of the
dilation operator. This information is encoded in the partition
function of the WZNW model. The latter decomposes into a sum of
products of characters for representations of the left- and right
moving chiral algebra. This does not mean that these models
experience holomorphic factorization -- they do not. But the
trace we take when we compute the partition function cannot see
the intricate coupling between left and right movers.

The representations of the affine $\widehat{\alg{psl}}(2|2)_{k}$
algebra along with their characters have been worked out for arbitrary
level $k$ in \cite{Gotz:2005ka}. When $k=1$, the theory contains a single
sector which is based on the vacuum representation of the current algebra.
Using the results of \cite{Gotz:2005ka} one can obtain the branching
functions for the decomposition of the affine modules into irreducible
representations of the zero-mode subalgebra $\psltt$. In case of the
vacuum representation of the affine $\psltt$ at level $k=1$ the branching
functions into representations $(j,l)$ of the even subalgebra $\g^{(0)}$
read
\begin{equation}
	\begin{split}
		\psi_{(j,l)}^{(0)}=\frac{q^{\frac{1}{12}}}{\phi(q)^{4}}
		\sum_{s\in\mathbb{Z}}\sum_{m,n=0}^{\infty}
		&\left( -1
		\right)^{m+n}q^{\frac{m(m+1)+n(n+1)}{2}+s(s+m-n)-j(m+n+1)}
		\\
		&\times \Bigl( 1 - q^{-(m+n+1)} \Bigr)\Bigl( 1 - q^{2l+1}
		\Bigr)q^{l^{2}},
	\end{split}
	\label{eqn:bos_branching_func}
\end{equation}
where $j,l\in\mathbb{Z}$ and with $s\to s+1/2$ for $j,l\in\mathbb{Z}+1/2$.
From these formulas one can determine the branching functions into representions
of the superalgebra $\g$ with the help of eqs.\ \eqref{eqn:atypical_rep_restriction}
and \eqref{eqn:typical_rep_restriction}. For the first few levels, the resulting
decomposition of the vacuum character $\hat \chi_0$ reads
\begin{equation}
	\begin{split}
		\hat \chi_0(q,x,y)&=q^{\frac{1}{12}}\bigl(
		q^{0}\chi_{[0]} + q^{1}\chi_{[\half]} +
		q^{2}(\chi_{[1,0]}+\chi_{[\half]}+\chi_{[0]})
		\\
		&+ q^{3}(\chi_{[2,0]}+\chi_{[1,0]}+ 2\chi_{[1]}+
		3\chi_{[\half]}+ 4\chi_{[0]})
		\\
		&+ q^{4}(\chi_{[3,0]}+ \chi_{[2,0]}+ 3\chi_{[1,0]}+
		\chi_{[0,1]}+ 2\chi_{[\frac{3}{2},\half]}
		\\
		&+ 2\chi_{[1]}+ 4\chi_{[\half]}+ 5\chi_{[0]})\bigr)
		+\mathcal{O}(q^{5})\ .
	\end{split}
	\label{eqn:partition_func_expansion}
\end{equation}
Here, we expanded the vacuum character of the affine $\psltt$ at level $k=1$
into characters $\chi_\Lambda = \chi_\Lambda(x,y)$ of the zero mode algebra
$\psltt$. The arguments $x,y$ keep track of the $\psltt$ weights while $q$
is associated with the eigenvalues of $L_0$, i.e.\ with the conformal weight
$h$, as usual. In the partition function, $\hat\chi_0$ gets multiplied with
an identical contribution from the anti-holomorphic sector, only that we need
to replace $q$ by $\bar q$.

From \eqref{eqn:bos_branching_func} it follows that the smallest conformal
weight at which a bosonic module $(j,l)\neq(0,0)$ appears is given by
\begin{equation}
	h^{\text{min}}_{g=0}(j,l)=\left\{
		\begin{aligned}
			&j+l^{2} & &\qquad j,l\in\mathbb{Z}
			\\[2mm]
			&j+l^{2}+\frac{1}{4} & &\qquad j,l\in\mathbb{Z}+\thalf.
		\end{aligned}
	\right.
	\label{eqn:bos_min_energy}
\end{equation}
In addition we note that modules with $j,l\in\mathbb{Z}+\half$
always appear with multiplicity two at their lowest weight. From the
decomposition \eqref{eqn:typical_rep_restriction} of typical
irreducible modules we can now deduce that the minimal weight
$h^{\text{min}}_{g=0}([j,l])$ of a module $[j,l]$, $j\neq l$, is given
by the minimal weight $h^{\text{min}}_{g=0}(j,l+1)$ of the bosonic
module $(j,l+1)$,
\begin{equation}
	h^{\text{min}}_{g=0}\bigl([j,l]\bigr)=\left\{
		\begin{aligned}
			&j+(l+1)^{2} & &\qquad j,l\in\mathbb{Z}
			\\
			&j+(l+1)^{2}+\frac{1}{4} & &\qquad
			j,l\in\mathbb{Z}+\thalf,
		\end{aligned}
	\right.
	\label{eqn:irred_min_energy}
\end{equation}
for typical $[j,l]$. With the help of the decomposition \eqref{eqn:atypical_rep_restriction} one can find a similar
result for atypical representations,
\begin{equation}
	h^{\text{min}}_{g=0}\bigl([j]\bigr)=\left\{
		\begin{aligned}
			&j^{2} + 2j & &\qquad j\in\mathbb{Z}
			\\
			&j^{2} + 2j -\frac{1}{4} & &\qquad
			j\in\mathbb{Z}+\thalf.
		\end{aligned}
	\right.
	\label{eqn:irred_atyp_min_energy}
\end{equation}
Given the values  \eqref{eqn:casimir} of the quadratic Casimir it is
clear that if we take $g\leq0$, high-gradient operators become relevant
for arbitrarily small values of the coupling, since their engineering
dimension grows linearly in $j$, while the anomalous dimension grows
like $-j^{2}$. So this direction of the perturbation cannot lead to a
stable theory.

Let us therefore turn to the case $g\geq0$. From \cite{Gotz:2005ka} we
know that operators that are invariant under the diagonal action of
$\psltt$ must transform in the same representation $\Lambda_L =
\Lambda_R$ with respect to the left and right action. Eq.\
\eqref{eqn:anom_dim_psl} implies that the only invariant operators
that become more relevant as we increase the coupling $g$ must sit in
multiplets
$\Lambda_L = [j,l] = \Lambda_R$ with $l > j$. Among those, the lowest
lying ones at $g=0$, namely those with $j=0$, are also those that
receive the largest correction to their conformal weights. From
eq.\ \eqref{eqn:anom_dim_psl}, the anomalous dimension
$\delta_g\bigl([0,l]\bigr)$ of invariant operators with $\Lambda_L =
[0,l] = \Lambda_R$ is given by
\begin{equation}
	\delta_g\bigl([0,l]\bigr)=-\frac{g}{1+g}l(l+1).
	\label{eqn:max_anom_dim}
\end{equation}
Comparing with eq. \eqref{eqn:irred_min_energy}, we infer that
these operators remain irrelevant for all finite values of the
coupling. In conclusion, the models with $g \geq 0$ actually
contain no RG-relevant invariant operators. Thereby, we have
extended the 1-loop result of \cite{Ryu:2010iq} to all loops
and all invariant operators.

\section{The spectrum at infinite coupling}
\label{sec:spec_at_inf_g}

The limiting point $g=\infty$ is obviously of special interest. Let us
therefore describe its spectrum in some more detail. From the above discussion
we can conclude that there are no relevant invariant operators in the spectrum
for any positive value of the coupling. Moreover, we see that as $g\to\infty$
the spectrum of operators in atypical ($\half$BPS) representations under the
diagonal action of $\psltt$ is half-integer valued, i.e.
\begin{equation}
h_{g} := h_{g=0} +  \delta_{g} \quad \mbox{satisfies} \quad
h_\infty = \lim_{g\rightarrow \infty} h_g  \in \tfrac12 \mathbb{Z}\ .
\end{equation}
The multiplicities of $\half$BPS states at any total conformal weight
$\Delta=h+\bar{h}$ remain finite as the coupling $g$ tends to infinity,
as can be seen with the help of  eq.\ \eqref{eqn:irred_min_energy} together
with eq.\ \eqref{eqn:anom_dim_psl}. For $j_{1}, j_{2} \in\mathbb{Z}$ we find
\begin{equation}
	\Delta^\text{min}_{\infty}
	=h^\text{min}_{\infty}\bigl([j_{1},l_{1}]\bigr)
	+\bar{h}^\text{min}_{\infty}\bigl([j_{2},l_{2}]\bigr) =
	j_{1}^{2}+2j_{1}+j_{2}^{2}+2j_{2}+l_{1}+l_{2}+2.
	\label{eqn:total_energy_at_infty}
\end{equation}
When either $j_{1}$ or $j_{2}$ are half-integer, $\tfrac{1}{4}$ gets added to
the above formula. If they are both half-integer, we must add $\tfrac{1}{2}$.
Since all the labels are non-negetive, the total energy grows strictly
monotonically in them. Therefore, multiplicities of $\half$BPS states
remain finite for any given value of $\Delta_{\infty}$. Moreover,
$\Delta_{\infty}$ remains non-negative and the only state that goes
to $\Delta_{\infty}=0$ is the ground state of the WZNW model.

We will now describe the spectrum at $g=\infty$ up to $\Delta_{\infty}=5$.
The analysis is organized according to the right moving conformal weight
$\bar{h}_\infty$, i.e. we shall start by listing all the $\half$BPS states
that possess $\bar{h}_\infty = 0$, i.e.\ the chiral states of the Gross-Neveu
model at strong coupling $g=\infty$. Obviously, all chiral $\half$BPS
states of the WZNW model, that is those that are built with the right
moving vacuum state and hence have weights $(h_{0},0)$, do not acquire an
anomalous contribution to their conformal weights. Hence, chiral states
of the WZNW model give states with $(h_\infty = h_0,\bar h_\infty = 0)$.
That does not mean,
however, that the chiral $\half$BPS spectrum at $g=\infty$ is the
same as it is at $g=0$. Indeed, starting from $h_{\infty}=3$ we see
new chiral states appearing. The first ones originate from an operator
multiplet at $\bigl(h_{0},\bar{h}_{0}\bigr)~=~(4,1)$ that transforms in
the representation $[0,1]^L\otimes[\half]^R$ in the WZNW model.
Under the diagonal action $D$, this product decomposes into
\begin{equation}
	[0,1]\otimes [\thalf] =
	6 [0]+6 [\thalf]+4 [1]+[\tfrac{3}{2}] + \text{typicals}.
	\label{eqn:tensor_prod}
\end{equation}
Hence, this multiplet of the WZNW model contributes plenty of chiral
fields at strong coupling. At $h_{\infty}=4$ we only need to account for the
holomorphic derivative of this operator. For $h_{\infty}=5$, finally,
there exist three multiplets in the representation $[0,1]^L\otimes [\half]^R$.
Additionally, we obtain a contribution from a multiplet that transforms in
$[0,2]^L\otimes[0,1]^R$. Its $\half$BPS
content in the decomposition with respect to the diagonal action is
the same as for the previous operator. Summing everything up, the
chiral spectrum to this level is given by
\begin{equation}
	\begin{aligned}
		h_\infty&=0 & &\qquad[0]
		\\
		h_\infty&=1 & &\qquad[\thalf]
		\\
		h_\infty&=2 & &\qquad[0] + [\thalf]
		\\
		h_\infty&=3 & &\qquad10 [0]+9 [\thalf]+6 [1]+[\tfrac{3}{2}]
		\\
		h_\infty&=4 & &\qquad11 [0]+10 [\thalf]+6 [1]+[\tfrac{3}{2}]
		\\
		h_\infty&=5 & &\qquad38 [0]+37 [\thalf]+24 [1]+5[\tfrac{3}{2}]\ .
	\end{aligned}
	\label{eqn:chiral_spec}
\end{equation}
The analysis for the next cases with $\bar{h}_\infty >0$ proceeds along
the same lines. For $\bar{h}_{\infty}=1$ one finds,
\begin{equation}
	\begin{aligned}
		h_\infty&=1 & &\qquad4 [0]+2 [\thalf]+2 [1]
		\\
		h_\infty&=2 & &\qquad4 [0]+3 [\thalf]+2 [1]
		\\
		h_\infty&=3 & &\qquad18 [0]+20 [\thalf]+14 [1]+5[\tfrac{3}{2}]
		\\
		h_\infty&=4 & &\qquad22 [0]+23 [\thalf]+16 [1]+5[\tfrac{3}{2}]\ .
	\end{aligned}
	\label{eqn:hbar_1}
\end{equation}
Similarly, the results for $\bar{h}_{\infty}=2$ read
\begin{equation}
	\begin{aligned}
		h_\infty&=2 & &\qquad19 [0]+16 [\thalf]+8 [1]+[\tfrac{3}{2}]
		\\
		h_\infty&=3 & &\qquad58 [0]+61 [\thalf]+46
		[1]+17[\tfrac{3}{2}]+2[2].
	\end{aligned}
	\label{eqn:hbar_2}
\end{equation}
For higher values of $\bar{h}_\infty\leq 5$ the multiplicities of
$\half$BPS multiplicities in the $g=\infty$ Gross-Neveu model can
be inferred from the list we provided, exploiting that the spectrum
is certainly symmetric under the exchange of left- and right movers.

There exist actually a few more states at $\Delta_\infty =5$ that
we have not listed yet. In fact, $\Delta_{\infty}=5$ marks the
first level at which states with negative left moving weight
$h_{\infty}<0$ appear in the spectrum. At the same time,
$\Delta_\infty =5$ is also the lowest value of the scaling weight
at which half-integer conformal weights $(h_\infty,\bar{h}_\infty)$
are actually observed.  The additional states are generated by two
WZNW operators that transform in the representation $[\half]^L\otimes
[\half,\frac{3}{2}]^R$. In this case, the decomposition of the
diagonal action can be worked out to give
\begin{equation}
	2[\thalf]\otimes [\thalf,\tfrac{3}{2}] = 4 [0]+8 [\thalf]+12
	[1]+8 [\tfrac{3}{2}]+2[2] + \text{typicals}.
	\label{eqn:tensor_prod_2}
\end{equation}
Note that we multiplied the left hand side by a factor $2$ so that
the left hand side accounts for all operators that possess weights
$\bigl(h_{\infty},\bar{h}_{\infty}\bigr) = \bigl(-\half,
\frac{11}{2}\bigr)$ at $g=\infty$. Of course, the spectrum is
symmetric under the exchange of the holomorphic and anti-holomorphic
sectors so that the same content appears with $\bigl(h_{\infty},
\bar{h}_{\infty} \bigr) = \bigl(\frac{11}{2},-\half\bigr)$.

\section{Conclusions and open problems}
\label{sec:conclusions}
Using exact results on the anomalous dimensions of operators in
perturbed WZNW models, we were able to show analytically that the $\psltt$
WZNW model does not contain RG-relevant high-gradient operators,
confirming the findings of \cite{Ryu:2010iq} and extending them
to all orders in the coupling and all invariant operators of the
model. This shows that the
$\pslNN$ WZNW models, at least for $N=2$, take a special role among
models with target-space supergroup symmetry. There is only one other case,
namely that of the boundary $\ospft$ Gross-Neveu model, in which
similar stability statements have been established \cite{Mitev:2008yt},
at least against boundary perturbations. Bulk perturbations, on the other
hand, were recently seen to produce strongly RG-relevant operators, much
in the same way as for other sigma models \cite{Cagnazzo:2014yha}.

We also observed that, as the coupling $g$ tends to plus infinity, the
spectrum becomes half-integer valued, indicating that the theory could
possess a free-field description. For the boundary $\ospft$\ Gross-Neveu
model, a similar study has been performed and the resulting spectrum has
been identified with a boundary spectrum of the free sigma model on the
supersphere $S^{3|2}$, see \cite{Mitev:2008yt}. It has been argued
several times before that the $\psltt$\ Gross-Neveu model should be
dual to the sigma model on $\CPot$\ \cite{CPdual}. At zero sigma model
coupling (infinite radius), the spectrum of boundary operators in that
model has been worked out in \cite{Candu:2009ep}. It is not difficult to
extend that analysis to the bulk, but unfortunately, the resulting spectrum
bears no resemblance with what we saw in the previous section. In any case,
it would be very interesting to identify a free field theory that gives
rise to the spectrum of the $\psltt$\ model at $g=\infty$ and possibly
to understand its precise relation to the $\CPot$ model.

\section*{Acknowledgments}
The authors wish to thank Constantin Candu, Vladimir Mitev, Andreas Ludwig,
Christopher Mudry, Thomas Quella and Hubert Saleur for comments and
interesting discussions. The research leading to these results has received
funding from the People Programme (Marie Curie Actions) of the European
Union's Seventh Framework Programme FP7/2007-2013/ under REA Grant
Agreement No 317089 (GATIS).


\end{document}